\documentclass[useAMS,usenatbib]{mn2e}

\usepackage{graphicx}
\usepackage{times}%
\usepackage{natbib}
\renewcommand{\d}{\mathrm{d}}
\newcommand{\gtrsim}{\ga}
\newcommand{\lesssim}{\la} 
\title[Early structure formation in quintessence models]
{Early structure formation in quintessence models and its implications
for cosmic reionisation from first stars}
\author[U. Maio et al.]
{U. Maio$^{1,2}$, 
K. Dolag$^{1}$,
M. Meneghetti$^{3,4}$,
L. Moscardini$^{2,5}$, 
N. Yoshida$^{6}$,
C. Baccigalupi$^{3,7,8}$,
\newauthor
M. Bartelmann$^{3}$,
F. Perrotta$^{3,7,8}$\\
$^1$ Max-Planck Institut fuer Astrophysik,
Karl-Schwarzschild Strasse 1, D-85748 Garching, Germany
(maio,kdolag@mpa-garching.mpg.de)\\
$^2$ Dipartimento di Astronomia, Universit\`a di Bologna,
via Ranzani 1, I-40127 Bologna, Italy
(lauro.moscardini@unibo.it)\\
$^{3}$ Zentrum f\"ur Astronomie, ITA, Universit\"at Heidelberg,
Albert-\"Uberle-Stra{\ss}e 2, D-69120 Heidelberg, Germany
(meneghetti,mbartelmann@ita.uni-heidelberg.de)\\
$^4$ INAF, Osservatorio Astronomico di Bologna, 
via Ranzani 1, I-40127 Bologna, Italy\\
$^{5}$ INFN, Sezione di Bologna, viale Berti Pichat 6/2,
I-40127 Bologna, Italy\\
$^{6}$ Department of Physics, Nagoya University, Nagoya, Aichi 464-8602, 
Japan (nyoshida@a.phys.nagoya-u.ac.jp)\\
$^{7}$ SISSA/ISAS, via Beirut 4, I-34014 Trieste, Italy 
(bacci,perrotta@sissa.it)\\
$^{8}$ INFN, Sezione di Trieste,
via Valerio 2, I-34127 Trieste, Italy\\
}

\begin{document}

\date{Accepted ???. Received ???; in original form July 2006}

\pagerange{\pageref{firstpage}--\pageref{lastpage}} \pubyear{2006}

\maketitle

\label{firstpage}

\begin{abstract}
We present the first hydrodynamic N-body simulations of primordial gas
clouds responsible for the reionisation process in dark energy
cosmologies.  We compare the cosmological constant scenario with a
SUGRA quintessence model with marked dynamics in order to highlight
effects due to the different acceleration histories imposed by the
dark energy.  We show that both the number density of gas clouds and
their clumpiness keep a record of the expansion rate during evolution,
similar to the non-linear dark matter profile at virialisation, as was
recently demonstrated by Dolag et al. (2004). Varying the shape of the
primordial power spectrum, we show how this effect is mitigated by a
running spectral index decreasing the power at small scales. Our
results demonstrate that, in order to constrain the dark energy from
large scale structures, one must track its effects down to the
distribution of luminous matter.
\end{abstract}


%*****************************************************************************

\section{Introduction} \label{sect:intro}

Different observational data, like those from high-redshift supernovae
\citep{riess2004,astier2006}, the cosmic microwave background
\citep{spergel2003,spergel2006}, and large-scale structure
\citep{tegmark2004,cole2005}, are now giving a consistent picture of
our universe, which can be described by the so-called `concordance'
$\Lambda$CDM model: a spatially flat universe whose expansion
accelerates in the present epoch because of a dominant dark energy
component.  Historically, the first and simplest candidate for that
has been the cosmological constant $\Lambda$, corresponding to matter
with an equation of state $p=w \rho c^2$, with constant $w=-1$.
However, observations suggest a value for $\Lambda$ which is more than
a hundred orders of magnitude smaller than the energy scales expected
to be responsible in the very early universe.

To avoid inelegant solutions based on parameter fine-tuning, the
general idea of dark energy is extended to encompass the so-called
quintessence, possibly corresponding to a suitably self-interacting
scalar field, whose pressure and energy density evolve during cosmic
history.  The currently available observational data sets do not allow
yet to place strong constraints on $w$. We know that it has to come
close to $-1$ with a precision of about 10 per cent at the present
epoch \citep[see, e.g.,][]{riess2004,spergel2006}, but its redshift
evolution is still poorly constrained and can hopefully be determined
only with new-generation data; see \citet{seljak2005} for one of the
very first attempts to measure the high-redshift dark energy equation
of state.

One consequence of $w>-1$ earlier is that the formation of cosmic
structures sets in earlier compared to the $\Lambda$CDM model. At a
given redshift, this corresponds to a higher abundance of more
concentrated halos \citep[see, e.g.,][]{dolag2004}. This may open
alternative ways to study quintessence models based on galaxy-cluster
counts \citep{wang1998,haiman2001,battye2003,majumdar2003} and strong
gravitational lensing \citep{bartelmann2003,meneghetti2005}.

The different history of structure formation must also affect the
reionisation epoch. To produce the required ionising radiation, a
sufficiently high number of early stellar sources is needed, such as
super-massive Pop-III stars or more ``standard'', massive Pop-II stars
in proto-galaxies. In particular, Pop-III stars are supposed to be
composed mainly of hydrogen and helium with their primordial
abundances, to have masses $\gtrsim100\,M_\odot$ much larger than the
standard Pop-I and -II stars, and to form in dark matter haloes with
typical masses of $10^{5\ldots6}\,M_\odot$ (the so-called mini-haloes).

The recent analysis of the temperature-polarisation cross-correlation
and the polarisation auto-correlation measured by the WMAP satellite
now allows the estimation of the epoch when reionisation occurs. The
Thomson optical depth of of $\tau\sim 0.17\pm 0.04$ extracted from the
first-year WMAP data had been interpreted as signalling the high
redshift of $z=17\pm5$ for the primordial epoch of global reionisation
\citep{kogut2003}. The recently released three-year WMAP data allowed
an improved control in particular of the polarised foreground
emission, yielding $\tau=0.09\pm0.03$. This corresponds to
$z=10^{+2.7}_{-2.3}$ for the completion of the reionisation process,
even if some level of parameter degeneracy still remains \citep[see,
e.g., Fig. 3 in][]{spergel2006}.\footnote{We have also to
notice that the  three-year WMAP data are suggesting a lower value
not only for $\tau$, but also for the power spectrum normalization
$\sigma_8$: these two effects nearly cancel in terms of early
structure formation, i.e. the inferred small $\tau$ does not really
allow slow reionization for a given $\sigma_8$.}  Different authors
\citep[see, e.g.,][]{wyithe2003,ciardi2003,sokasian2003,sokasian2004}
suggested that these reionization data can place new complementary
constraints on the cosmological parameters, and in particular on the
nature of dark energy.

So far, studies of the structure formation process in dark energy
cosmologies were focused on modifications of dark matter structures
due to the different expansion histories \citep[see,
e.g.,][]{klypin2003,dolag2004}. They revealed that virialised objects
keep a record of the expansion rate at the time of their
formation. The general picture is that in cosmologies with $w>-1$, the
increase of the dark energy density with redshift enhances structure
growth, and virialised haloes become more concentrated because of the
denser environment they form in \citep{dolag2004}. This effect is
stronger in tracking quintessence models, which we describe in the
next section, in which the increase of the dark energy density with
redshift is enhanced. Complementary hydrodynamical simulations still
need to be carried out. This work is a first step into this direction,
beginning with the numerical study of the dependence of the primeval
gas clouds responsible for the reionisation process on the cosmic
expansion rate at their formation time.

Due to the higher concentration of structures in dark energy
scenarios, which has so far been verified only for the dark matter, it
is mandatory for our study to control the shape of the primordial
power spectrum. When combined with the Ly-$\alpha$ forest data and
with the analysis of the $2dF$ galaxy redshift survey, the first-year
WMAP results supported a ``running'' spectral index which tilts the
spectrum slightly towards small scales, starting at
$k\gtrsim1\,\mathrm{Mpc^{-1}}$. The three-years WMAP data are
compatible with this \citep{spergel2006}, albeit with less emphasis
than the earlier results. A running index would cause a slower growth
and evolution of small-scale cosmic structures compared to models with
a constant index $n$. A reduction of power on small scales may
alleviate several potential discrepancies in the $\Lambda$CDM model,
such as the abundance of substructures in galaxies and the high
central concentration of galactic haloes. However, as shown by
analytic and numerical work \citep[see,
e.g.][]{somerville2003,yoshida2003b}, models with a running spectral
index (RSI) may have severe problems in producing enough objects to
allow a global reionisation at high redshift. Since this might be
balanced by the enhanced structure growth due to the dark energy, it
is important to jointly study these two aspects.

In the present paper, we study the high-redshift structure formation
in quintessence models based on the results of high-resolution
cosmological N-body simulations combined with
hydrodynamics. Specifically, we consider four different cases
combining two flat cosmological models, i.e.\ the concordance
$\Lambda$CDM model and a quintessence model with a SUGRA potential,
with two types of the primordial power spectrum, one with a constant
spectral index $n=1$ and an RSI model assuming the best fit relation
found by \cite{spergel2003}.  More detail will be given in the next
section. The paper is organised as follows. In Sect.~\ref{sect:sim},
we describe the general characteristics of the hydrodynamical
simulations used below and introduce the quintessence models
chosen. The techniques adopted for identifying dark matter haloes and
the corresponding results are presented in
Sect.~\ref{sect:haloes}. The abundances of gas clouds are presented in
Sect.~\ref{sect:clouds} together with the clumping factors and
recombination times as computed from the simulation outputs. The
implications of the previous results in terms of reionisation are
discussed in Sect.~\ref{sect:reionization}. The final discussion and
our main conclusions are drawn in Sect.~\ref{sect:conclusions}.

%*****************************************************************************

\section{The simulations} \label{sect:sim}

We first review the cosmologies chosen, and then the computational and
hydrodynamical features of our numerical simulations. As mentioned in
the introduction, this work is a first step towards investigating the
dependence of  hydrodynamic N-body simulated structures on the
global cosmic expansion rate, and it is appropriate to study the
connection with the shape of the primordial power spectrum. Thus, we
need dark energy cosmologies with different dynamics as well as
constant and running index in the primordial power spectrum.

The standard, concordance $\Lambda$CDM model is our reference case. It
is spatially flat, has the present matter density $\Omega_{\rm
0m}=0.3$, the baryon density $\Omega_{\rm 0b}=0.04$, and a dominating
cosmological constant, $\Omega_{0\Lambda}=0.7$. The Hubble constant is
assumed to be $h=0.7$ in units of
$100\,\mathrm{km\,s^{-1}\,Mpc^{-1}}$. The second model is chosen so as
to emphasise the dynamical effects of the dark energy. The spatial
curvature is also vanishing, with $\Omega_{\rm 0m}=0.3$ and 70 per cent of the
critical density contributed by a quintessence scalar field $\Phi$
evolving in time under the effect of a potential energy $V(\Phi)$. We
review here the main characteristics of this model, referring to
\citet{peebles2003} and references therein for further detail.

The field $\Phi$ must satisfy the Klein-Gordon equation
\begin{equation}
\label{KG}
\ddot{\Phi}+3H\dot{\Phi}+\frac{\partial V}{\partial \Phi}=0\ .
\end{equation}
The equation-of-state parameter $w\equiv p/(\rho c^2)$ is
\begin{equation}
\label{PSW}
w=\frac{\frac{1}{2}\dot{\Phi}^2-V(\Phi)}{\frac{1}{2}\dot{\Phi}^2+V(\Phi)}\ .
\end{equation}
Evidently, $w$ evolves with redshift, but the behaviour of the
cosmological constant is reproduced as $w\rightarrow-1$ in the static
limit $\dot{\Phi}^2 \ll V(\Phi)$.

The attractive feature of such models are their attractor solutions
which exist for exponential and inverse power law potentials or
 combinations of those. Suitably tuning the potential amplitude, they can
reach the present level of dark energy starting from a large range of
initial conditions. The evolution of the equation of state may be more
or less pronounced along these trajectories. We wish to have
pronounced dark-energy dynamics in order to clearly identify its
effects. Thus, we choose a SUGRA potential
\begin{equation}
V(\Phi)=\frac{M^{4+\alpha}}{\Phi^{\alpha}}\exp\left(4\pi G\Phi^2
\right)\ ,
\end{equation}
 where G is the Newton constant, 
as suggested by super-gravity corrections
\citep{brax1999,brax2000}.  By exploiting the properties 
of tracking trajectories, we fix $M$ by requiring that the dark energy 
makes up 70 per cent of the critical density today, and $\alpha=6.5$ so 
that $w\to w_0\equiv w(z=0)=-0.85$ today. 
At higher redshift, $w>w_0$ \citep{brax2000}, which meets our purpose 
of emphasising differences with the $\Lambda$CDM model. It should be noted 
that this model is already at variance with the observational constraints 
\citep[see,e.g., ][]{spergel2003,knop2003,riess2004}.

\begin{figure} 
 \includegraphics[width=0.45\textwidth]{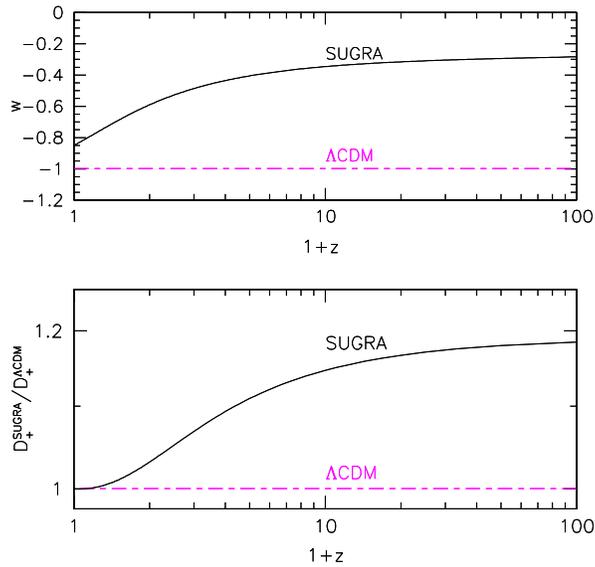}
\caption{\emph{Upper panel}: The redshift evolution of the
equation-of-state parameter $w$ for the SUGRA model with $w_0=-0.85$
(solid line), compared to the concordance $\Lambda$CDM model
(dash-dotted line). \emph{Lower panel}: The linear growth factor
(normalised to the $\Lambda$CDM model at present) for the same
cosmologies as in the upper panel.}
\label{fig:wz}
\end{figure}

The upper panel of Fig.~\ref{fig:wz} shows the redshift evolution of
$w$ for this model. The differences with the standard $\Lambda$CDM
model (constant $w=-1$) are large at all redshifts. In particular,
during the epoch of reionisation which we are mostly interested in,
the SUGRA model has $w\approx -0.35$. Notice that $w$ strongly affects
the formation history of cosmic structures through its modified
expansion history. This is evident from the lower panel in
Fig.~\ref{fig:wz}, which shows the linear growth factor (normalised to
coincide with the $\Lambda$CDM model at $z=0$) for the model shown
above.  In quintessence models where the dark energy is more abundant 
in the past with respect to now,  structures tend to form earlier 
if the perturbation power spectrum is normalised at the present epoch. 
This is due to the earlier dark energy dominance, slowing down 
the perturbation growth.
In our SUGRA model with $w_0=-0.85$, fluctuations at $z=20$ are expected 
to be $\approx15$ per cent higher than in the $\Lambda$CDM model. 
Consequently, we expect that non-linearity to be reached earlier, 
with obvious consequences for the reionisation epoch.

\begin{figure} 
 \includegraphics[width=0.45\textwidth]{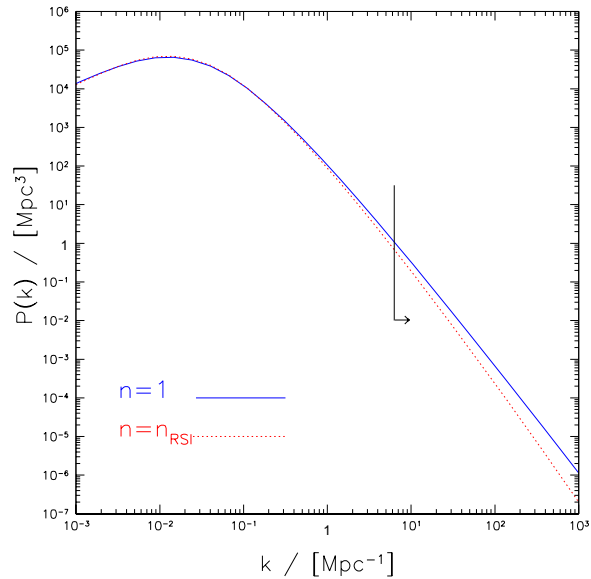}
\caption{Comparison between the two different CDM power spectra
adopted in generating the initial conditions for the numerical
simulations: solid and dotted lines show spectra with fixed $n=1$ and
with the running spectral index (\ref{FIT}), respectively. The arrow
indicates the range of length scales covered by our simulations.}
\label{fig:parameters}
\end{figure}

Dynamical dark energy generally increases the concentration of
virialised structures \citep{dolag2004}, which can also be caused by a
modified primordial power spectrum. Therefore, we vary the simulation
set-up using two different shapes of the primordial power
spectrum. Specifically, for both cosmological models ($\Lambda$CDM and
SUGRA), we construct initial conditions corresponding to two
scenarios, namely a power spectrum with the standard constant $n=1$
normalised to $\sigma_8=0.9$. The second model also has
$\sigma_8=0.9$, but allows for a reduced primordial power on small
scales as suggested by the combined analysis of the first-year WMAP
data, galaxy surveys and Ly-$\alpha$ data \citep[see,
e.g.][]{spergel2003}, within the range compatible with the three-year
WMAP data \citep{spergel2006}.

The case for a running spectral index is interestingly controversial.
Re-analyses of the Ly-$\alpha$ data \citep[see,
e.g.][]{viel2004,seljak2005,viel2006,zaroubi2006} lend much less
support for a running spectral index than reported based on the
first-year WMAP data and favour a constant primordial spectral index
very close to unity 
\citep[see, however,][]{desjacques2005}. On the other hand, 
a running spectral index is still viable and fits the three-year WMAP
data well
\citep{spergel2006}, while the support of a pure power law from large
scale structures persist \citep[see,
e.g.,][]{viel2006b,seljak2006}. As mentioned before, we choose our
cosmological models so as to cover a broad range of phenomenologies
rather than to fit the existing data well.  Thus, we choose to adopt a
primordial power spectrum given by $P(k)\propto k^{n(k)}$, with
\begin{equation}
\label{FIT}
n(k)=n(k_0)+\frac{1}{2} \frac{{\rm d}n}{{\rm d} \ln{\rm k}} 
\ln \left(\frac{k}{k_0}\right)\ ,
\end{equation}
where  $k_0=0.05$ Mpc$^{-1}$, $n(k_0)=0.93$ and
\begin{equation} 
\frac{{\rm d}n}{{\rm d} ln{\rm k}} = -0.03\ .
\end{equation}
This choice conveniently allows a direct comparison with the similar
analysis by \cite{yoshida2003b}, who adopt the same primordial power
spectrum, but only consider the concordance $\Lambda$CDM model. It is
worth recalling that this represents the best fit to the first-year
WMAP data and remains compatible with the three-year data, $\d n/\d
\ln k\lesssim-0.06$
\citep{spergel2003,spergel2006}. Figure~\ref{fig:parameters} shows the
two primordial power spectra considered here.  Evidently, they differ
at large $k$ where the RSI spectrum is steeper. Less power on small
scales will delay the structure formation \cite[see also the
corresponding discussion in][]{yoshida2003b}.

\begin{figure*} 
 \includegraphics[width=0.49\textwidth]{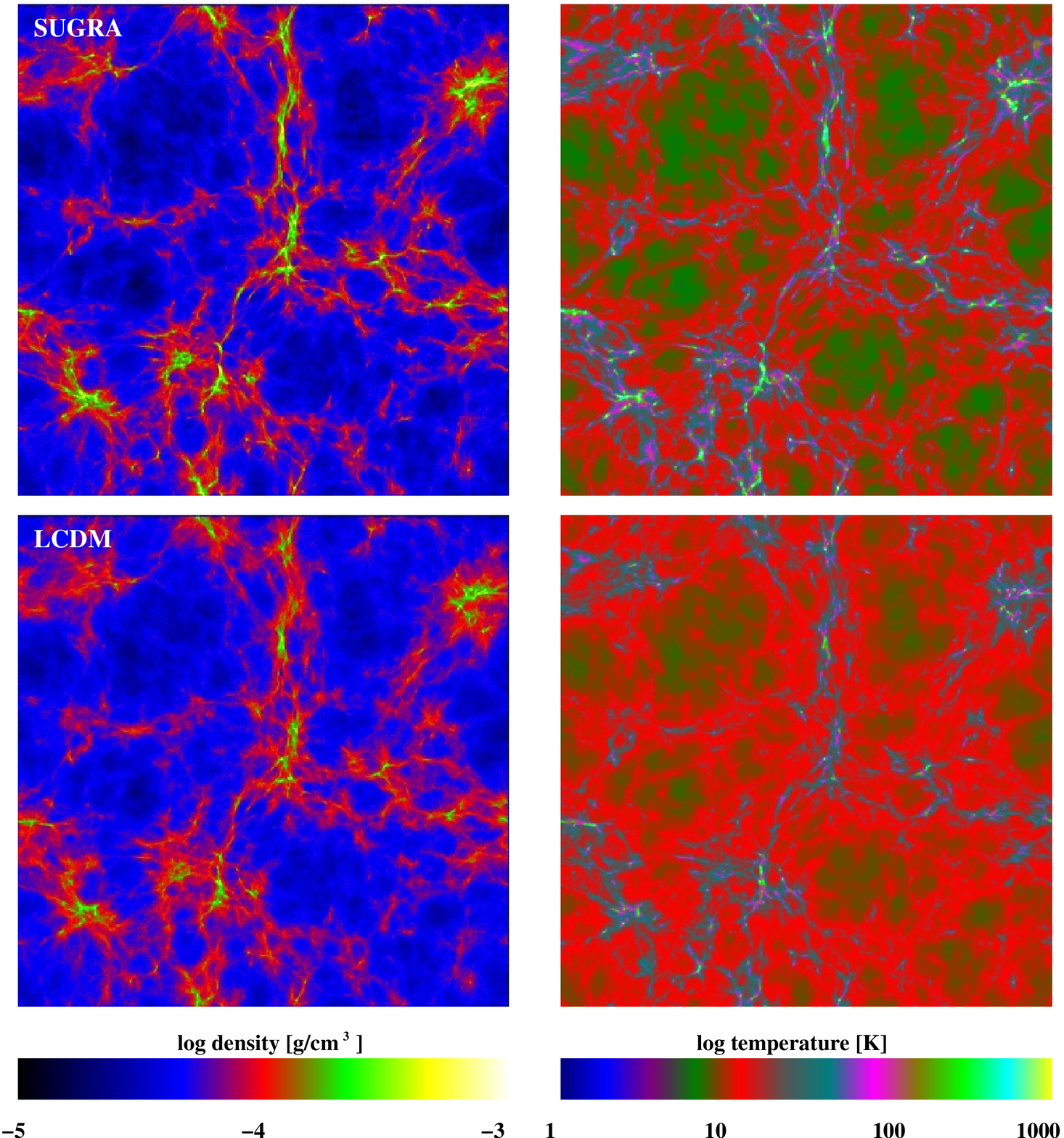}
 \includegraphics[width=0.49\textwidth]{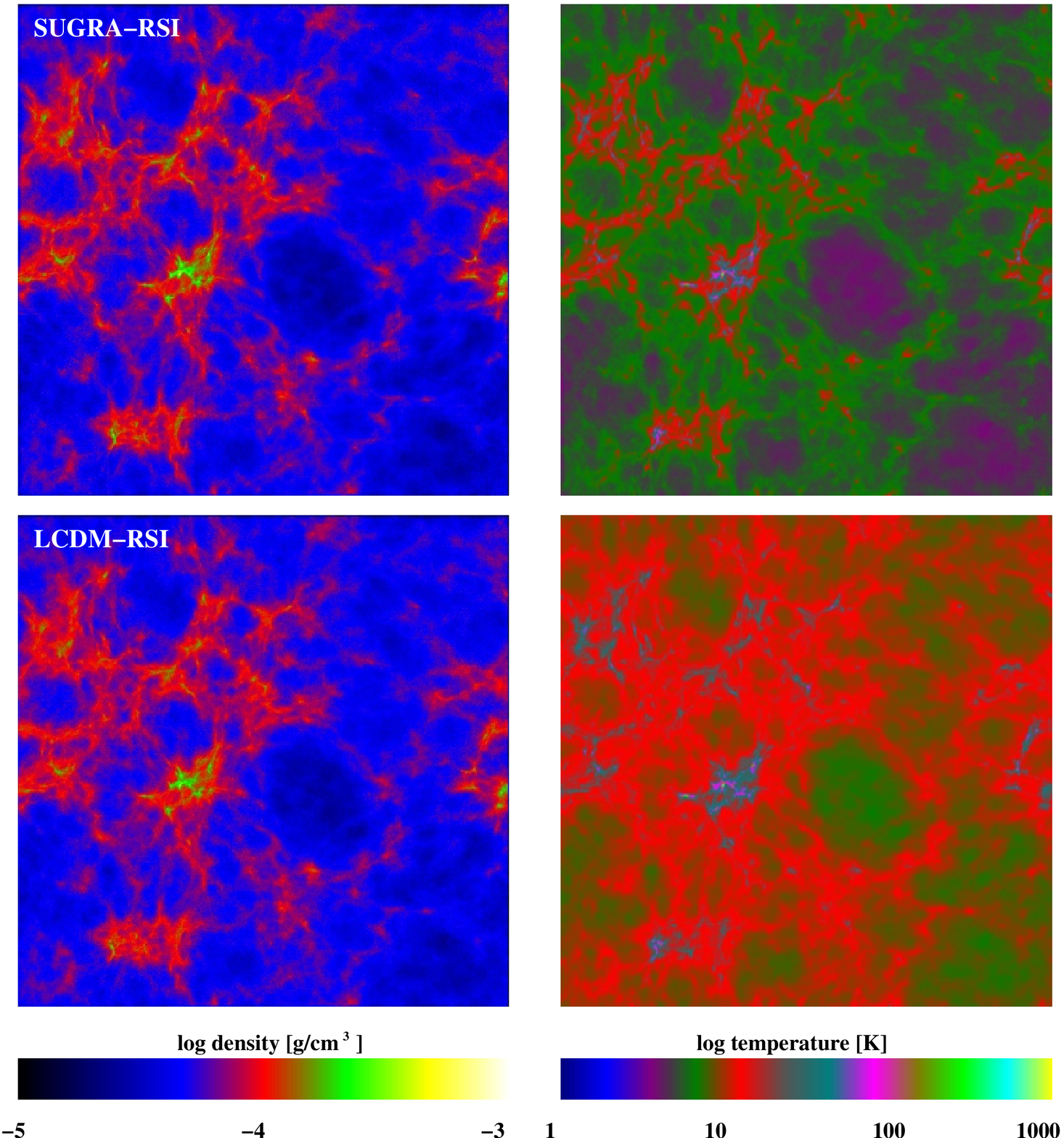}
\caption{The logarithm of the projected gas-density and mass-weighted
temperature at $z=20.6$ is shown in thin slices (1/7th of the box
height) for the different models: SUGRA, $\Lambda$CDM, SUGRA-RSI and
$\Lambda$CDM-RSI, as labelled in the maps. The density is in physical
units and the slices are 6.6 (physical) kpc thick at $z=20.6$. Note that the
models with and without RSI have different random phases in their
initial conditions and are thus not supposed to reproduce the same
structures.}
\label{fig:slices}
\end{figure*}

Combining two cosmological models ($\Lambda$CDM and SUGRA) with two
primordial power spectra (constant $n=1$ and RSI), we ran four
different simulations; $\Lambda$CDM and SUGRA with $n=1$ and
$\Lambda$CDM-RSI and SUGRA-RSI with a running spectral index. The
initial conditions were generated at $z\approx120$ using a comoving
periodic box of 1~Mpc side length. We followed \cite{dolag2004} in
adapting the initial conditions for $\Lambda$CDM to those for the
SUGRA model. Thus, by construction, only the pairs of models sharing
the same primordial power spectrum have the same random phases,
allowing a direct comparison of the structures formed. The models with
different primordial power spectra have different random phases and
can thus not be directly compared (see Fig.\ref{fig:slices}).

Since we are interested in structures during cosmic reionisation, the
box size used here can be considered sufficiently large to suppress
cosmic variance \citep[see the discussion on effects of finite box
sizes in][]{yoshida2003b, bagla2006}. Moreover, to prevent the
non-linear regime from reaching the fundamental fluctuation mode in
the simulations, we stopped the simulation at the latest at $z\sim15$.

When the gas starts cooling significantly within the first forming
structure, the simulation time step can become very small (since we
neglect feedback in these simulations), effectively stopping the
simulation. Accordingly, we had to stop some simulations at a higher
redshift, for example the SUGRA simulation, which could only be
evolved until $z\sim19$ because of its more rapid evolution, as we
shall see in the later analysis. Nonetheless, all simulations are
followed long enough for them to form the sufficient amount of
structure we need for our analysis.

The simulations were performed using the GADGET2 code
\citep{springel2001,springel2005} on the IBM-SP4/5 at CINECA
(Bologna).  GADGET2 is based on the combination of a
tree-particle-mesh algorithm to solve the gravitational forces, and
the smoothed particle hydrodynamics (SPH) scheme to describe the
hydrodynamics of gas particles. The code follows the non-equilibrium
reactions of different chemical species ($e^-$, $H$, $H^+$, $H^-$,
$He$, $He^+$, $He^{++}$, $H_2$, $H_2^+$), using the reaction
coefficients computed by \cite{abel1997} \citep[for more detail, see
also][]{anninos1997,yoshida2003a}, and adopts the cooling rate due to
molecular hydrogen estimated by \cite{galli1998}. The density field
was sampled with $324^3$ dark matter (DM) particles and an equal
number of gas particles, having a mass of $\sim1040\,M_\odot$ and
$160\,M_\odot$, respectively. The comoving Plummer-equivalent
gravitational softening length was fixed to
$\epsilon=0.143\,\mathrm{kpc}$.

In Fig.~\ref{fig:slices} we show the projected gas and (mass-weighted)
temperature distributions for all four models at $z=20.6$. As
expected, the RSI simulations appear delayed, exhibiting smoother
density and temperature distributions. The SUGRA models reveal their
expected behaviour with a more evolved density field than their
$\Lambda$CDM counterparts. Since the models with and without RSI have
different random phases in their initial conditions, they are not
supposed to reproduce the same structures.

\section{The abundance of dark matter haloes}\label{sect:haloes}

\begin{figure*}
\includegraphics[width=0.95\textwidth]{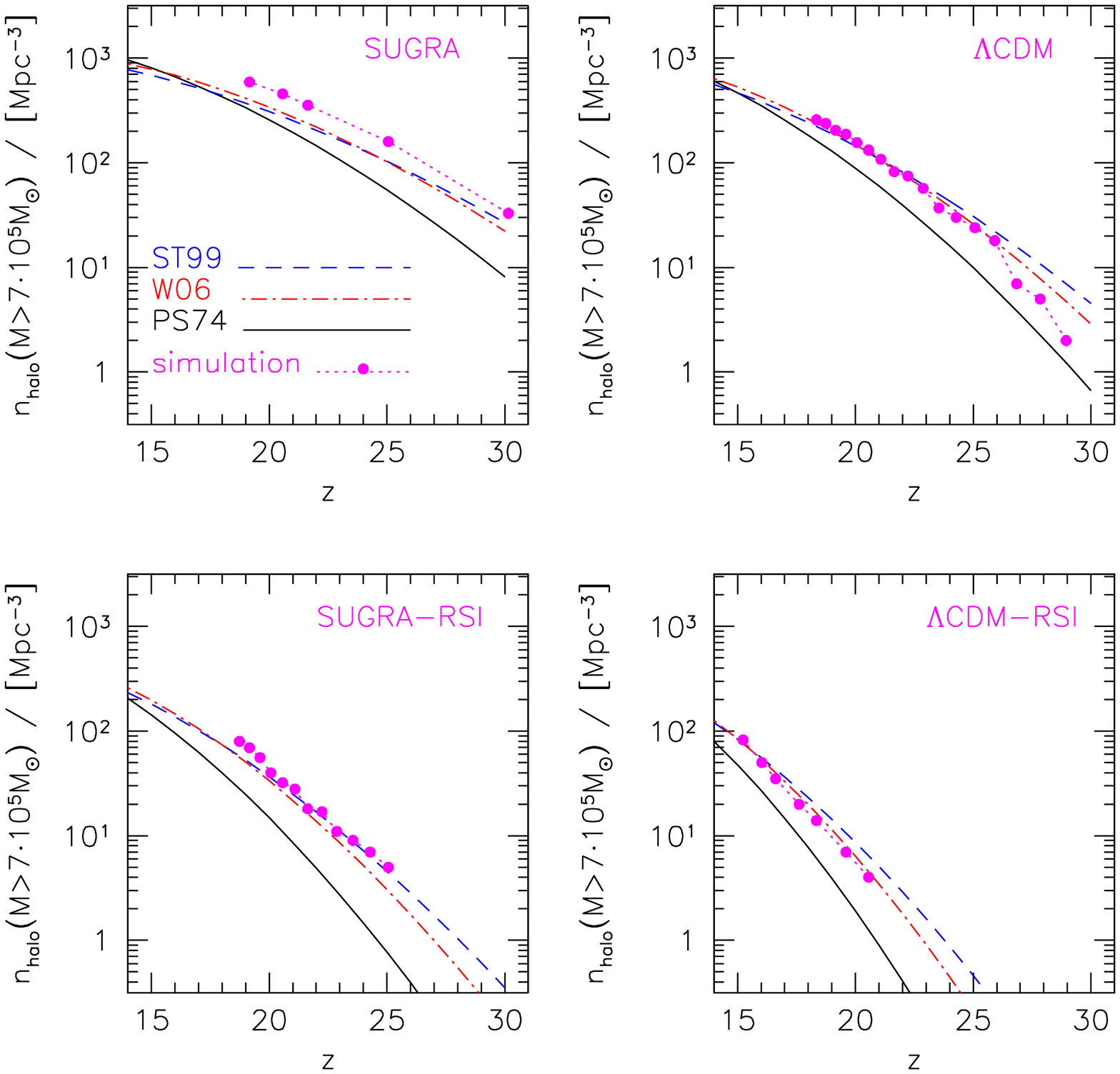} 
\caption{Number of haloes with mass exceeding $7\times10^5\,M_\odot$
per unit comoving volume as function of redshift in the different
simulations (filled circles and dotted lines): SUGRA (upper left
panel), $\Lambda$CDM (upper right panel), SUGRA-RSI (lower left
panel), $\Lambda$CDM-RSI (lower right panel). Theoretical predictions
based on the Press \& Schechter (PS74, solid line), Sheth \& Tormen
(ST99, dashed line) and the Warren et al. (W06, dot-dashed line)
relations are shown as well.}
\label{fig:Nhalo}
\end{figure*}

The abundance of virialised dark matter haloes is usually expressed by
the mass function $N(M)$. From a theoretical point of view, different
relations are available to predict the redshift evolution of $N(M)$
once the cosmological model is fixed. Some of these were obtained from
a specific model for the collapse of structures
\citep{press1974,sheth2002}; in other cases, they represent best fits
to N-body simulations with very high resolution
\citep{sheth1999,jenkins2001,warren2006}. However, the theoretical
mass functions have not yet been extensively tested in the regime of
very low masses and very high redshifts which we are mainly interested
in, where their predictions differ by orders of magnitude. This is due
to the difficulty of combining sufficient resolution in space and mass
with regions large enough to be considered a fair samples of our
universe.

The existing studies yield contradictory results. Simulations covering
large volumes, such as those presented in \cite{reed2003} ($432^3$
particles in a $50\,h^{-1}\,\mathrm{Mpc}$ box) and in
\cite{springel2005b} ($2160^3$ particles in a
$500\,h^{-1}\,\mathrm{Mpc}$ box), found that the \citep[][ hereafter
ST99]{sheth1999} formula fairly reproduces the numerical results, but
they can safely identify haloes with masses $\gtrsim10^{10}\,M_\odot$
at $z\lesssim 10$. At $z=15$, \cite{reed2003} found a slight tendency
of the ST99 model to over-predict the halo counts.

Numerical work based on much smaller boxes, such as
\cite{jang-condell2001} ($128^3$ particles in a
$1\,h^{-1}\,\mathrm{Mpc}$ box) or \cite{yoshida2003b,yoshida2003c}
($2\times 324^3$ particles in a $1\,h^{-1}\,\mathrm{Mpc}$ box), found
very good agreement with the predictions of the \citep[][ hereafter
PS74]{press1974} model. Notice that these simulations are more
affected by their finite box size (see the following
discussion). Further support for the PS74 formalism at high redshifts
is provided by the analysis of \cite{gao2005}, who showed that the
extended Press-Schechter theory \citep{bond1991} correctly predicts
the structure growth and the dependence of halo abundances on the
density of the surrounding region in the redshift range between 50 and
30. One of the best compromises between mass resolution and volume
coverage was achieved by \cite{heitmann2006}, who used an extended set
of simulations with $256^3$ particles in boxes sized between
$(4\ldots126)\,h^{-1}\,\mathrm{Mpc}$. Their mass function at high
redshift turns out to deviate significantly from the PS74 prediction
for $M>10^7\,h^{-1}\,M_\odot$, while the agreement with the \citep[][
hereafter W06]{warren2006} relation is better. Notice that the W06
formula predicts a substantial suppression of the halo abundance in
the high-mass regime compared to the ST99 model.  Finally, using an
array of high-resolution N-body simulations with box sizes in the
range $1-3000 h^{-1}$Mpc, Reed et al. (2006) recently determined the
mass function of dark matter haloes at redshifts between 10 and 30.
They found that the PS74 models gives a poor fit at all epochs, while
the ST99 model gives a better match to the simulations, but still
overpredicts the abundance of most massive objects by up to 50 per
cent.

We identify dark matter haloes in our simulations by running a
friends-of-friends algorithm on the dark matter particles only and
setting the linking length to 20 per cent of the mean inter-particle
separation. The resulting mass of each halo is then corrected by the
factor $(1040+160)/1040\approx1.15$ to account for the additional
contribution by the gas particles. Figure~\ref{fig:Nhalo} displays the
redshift evolution of the number of haloes with masses exceeding
$7\times10^5\,M_\odot$ for all four simulations. The figure compares
the results with the predictions of the PS74, the ST99 and the W06
models, indicated by solid, dashed and dot-dashed lines,
respectively. Note that we do not consider the relation found by
\cite{jenkins2001} because it cannot extrapolate to the mass and
redshift ranges considered here.

At first sight, the dark energy dynamics as well as the modified
initial power spectrum generate significant differences. First, as
expected, the number of massive haloes is always lower in the RSI
models than in their scale-invariant counterparts. While the first
objects with $M>7\times10^5\,M_\odot$ start appearing near $z\simeq30$
in the SUGRA and the $\Lambda$CDM simulations, this is delayed to
$z\simeq25$ and $z\simeq20$ for SUGRA-RSI and $\Lambda$CDM-RSI,
respectively.

Second, the different expansion histories leave clear records in the
structure population. Considering models with the same primordial
power spectrum, the halo abundance is always larger in the SUGRA than
in the $\Lambda$CDM models by a factor $\simeq2$, confirming the
earlier structure formation in dynamical quintessence models expected
from the modified linear growth factor in Fig.~\ref{fig:wz}.

We now consider the difference to the theoretical expectations from
the PS74, ST99 and W06 mass functions. All analytic relations were
modified so as to account for the different expansion histories and
initial power spectra in the different cosmological models. For the
two simulations with a cosmological constant ($\Lambda$CDM and
$\Lambda$CDM-RSI), the results indicate that the W06 formula agrees
better with the simulations than the ST99 and PS74 formulae. Some
small deviations are seen at very high redshifts where, however,
counts are low and the statistical uncertainty is high. The ST99 mass
function tends to slightly overestimate the simulation results, again
mainly at high redshift, where the discrepancies with W06 are more
evident. On the contrary, the PS74 mass function always underestimates
the halo abundances, with differences reaching up to a factor
$\sim3$. Thus, our results for the cosmological-constant models agree
with the previous analyses by \cite{reed2003}, \cite{springel2005b}
and \cite{heitmann2006}.  However, this conclusion is in conflict with
other studies \citep{jang-condell2001,yoshida2003b,yoshida2003c}.

The situation is different for both SUGRA models.  The mass function
in the simulation with constant $n=1$ is always larger than all
theoretical predictions. In particular, the ST99 and W06 relations
give very similar estimates which fall below the simulation results by
30 per cent, while the difference with PS74 is always a factor of 3-4. The
tendency of PS74 to underestimate the numerical mass function is also
confirmed for SUGRA-RSI (a factor of 3-5 everywhere), while the
agreement for W06 and mostly for ST99 is substantially better. The
deviations reach 20 per cent only at low redshift. These results show that
caution must be applied when applying the standard theoretical mass
functions to very high redshift in quintessence models.

We must finally remark that, owing to the much higher mass resolution,
the mass range covered in our simulations is quite different compared
to earlier work. At the same time, our results could be affected by
the small box size. In fact, as discussed in \cite{yoshida2003b}, the
``effective'' variance, i.e.~the amount of power on the scales covered
by the simulation, can be significantly smaller than that obtained by
integrating the power over all scales. This difference can
systematically delay structure formation and possibly cause the
discrepancies between the simulation results from different box
sizes. To quantify the effect of the finite box, we apply the analytic
expressions proposed by \cite{bagla2006}, which allow an evaluation of
the required correction to the mass function. For example, assuming
the PS74 approach and the redshift range covered by the data shown in
Fig.~\ref{fig:Nhalo}, we find that the correction for
$M=7\times10^5\,M_\odot$ leads to a slight underestimate of 10-15 per cent,
which does not significantly affect our previous results. Similar
conclusions can be reached on the ST99 relation.

\section{The properties of the gas distribution}\label{sect:clouds}

We now show our results on the statistics and the physical properties
of the gas distribution. We concentrate on two aspects, namely the
formation time of the gas clouds and their clumping factor, focusing
on their dependence on the underlying dark energy cosmology and
primordial power.

\subsection{The formation of gas clouds}\label{sect:num_clouds}

We first study the formation of the gas clouds which are responsible
for reionisation. For this purpose, we run a friends-of-friends
algorithm to group cold and dense gas particles, assuming a linking
length of 1/20th of the mean inter-particle separation to locate the
densest gas concentrations.  Following
\cite{yoshida2003b,yoshida2003c}, we only consider clouds exceeding
$3000\,M_\odot$, which corresponds to groups with at least 19 gas
particles in our simulations.

\begin{figure}
 \includegraphics[width=0.45\textwidth]{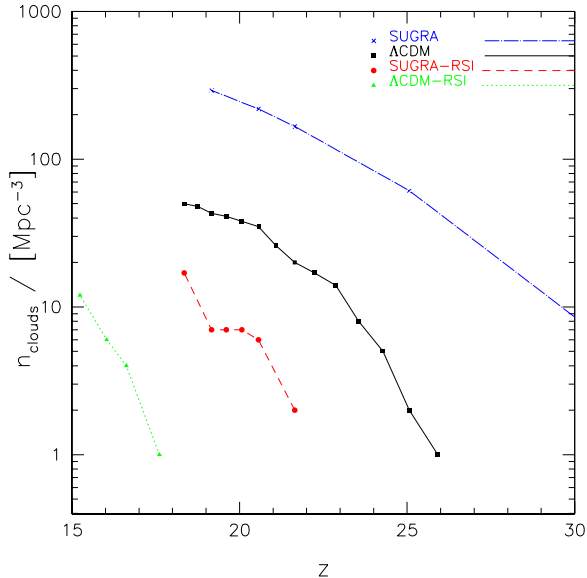} 
\caption{Number of clouds per unit comoving volume as function of
redshift for SUGRA, $\Lambda$CDM, SUGRA-RSI and $\Lambda$CDM-RSI, as
labelled.}
\label{fig:Nclouds}
\end{figure}

The results are shown in Fig.~\ref{fig:Nclouds} in terms of the
redshift evolution of the number of gas clouds $n_{\rm clouds}$
identified in the four simulations. It is not surprising that the trend
between the models is similar as for the dark matter haloes in
Fig.~\ref{fig:Nhalo}. Gas clouds start forming earlier and are more
abundant in the SUGRA than in the $\Lambda$CDM model with the same
primordial power spectrum.

The suppression of power on small scales is, however, the dominant
effect. In fact, the two RSI models have their structure formation
delayed and have the smallest number of gas clouds in the final
outputs. At $z\simeq19$, there are order-of-magnitude differences
between the models. We find $\sim300$ clouds for SUGRA, $\sim50$ for
$\Lambda$CDM, 7 for SUGRA-RSI and no clouds for
$\Lambda$CDM-RSI. Also, the formation of the first gas clouds strongly
depends on the cosmological parameters. It happens much before $z=30$
for SUGRA, while it is $z\simeq 18$ for $\Lambda$CDM-RSI. The reason
for this pronounced difference is the integral nature of the
observable $n_\mathrm{clouds}$ considered here. The production rate of
clouds per unit time is the key quantity, which differs between the
cosmologies considered because of the different expansion histories
and primordial power spectra. The SUGRA model produces clouds
earliest, followed by $\Lambda$CDM, while the same models with RSI
stay in the same order but the structure formation process is delayed
because of the modified primordial power.  The integrated cloud
count $n_\mathrm{clouds}$ accumulates these differences during the
whole epoch of cloud production.

Finally, we have to notice that in our study we neglected the feedback
effects which might influence the structure formation history of
primordial clumps.  For example, radiative feedbacks, such as
photoionisation or molecule photo-dissociation, could stop cooling
suppressing the formation of small objects; mechanical feedbacks and
shocks can determine gas removal from the structure, on one hand, or
induce further shell fragmentation, on the other hand.  Many studies
on this subject are present in the literature
\citep[see the corresponding discussion in ][and
references therein]{ciardi2005}, but up to now comprehensive
conclusions have not been reached: in fact the variety of approaches,
approximations and parametrizations adopted by different authors leads
to results which are often discordant.

\subsection{Clumping factor and recombination time}
\label{sect:C&Trec}

Following \cite{CLUMP}, we adopt a simple statistical description of
the reionisation process. Noting that each ionising photon emitted is
either absorbed by a newly ionised or by a recombining hydrogen atom,
the filling factor $Q(t)$ of regions of ionised hydrogen at any given
time $t$ follows from subtracting the total number of recombinations
per atom from the total number of ionising photons per hydrogen atom
emitted before $t$. In this way, it is possible to derive a simple
differential equation describing the transition from the neutral to
the completely reionised universe \citep{MADAU},
\begin{equation}
\frac{\d Q}{\d t}=
\frac{\dot n_{\rm ion}}{\bar n_H}-\frac{Q}{t_{\rm rec}}\ ,
\label{eq:dqdt}
\end{equation}
where $\bar n_H$ represents the mean comoving density of hydrogen
atoms and $\dot{n}_{\rm ion}$ is the comoving emission rate of photons
capable of ionising hydrogen. The recombination time of hydrogen,
$t_{\rm rec}$, is
\begin{equation}
t_{\rm rec}=\frac{1}{\alpha_B\, C\, {\bar n_H}\, (1+z)^3 \, (1+2\chi)}\,\,,
\label{eq:trec}
\end{equation}
where $\alpha_B$ the recombination coefficient and $\chi$ is the
relative abundance of helium with respect to hydrogen. The
dimensionless clumping factor $C$ of hydrogen is given by
\begin{equation}
C\equiv \frac{\langle n^2_{H^+}\rangle}{\langle n_{H^+} \rangle^2}\,\,.
\end{equation}
It is evident from Eq.~(\ref{eq:trec}) that the recombination time is
a function of redshift. Assuming a gas temperature of
$10^4\,\mathrm{K}$ above which line cooling by atomic hydrogen becomes
efficient, the previous relation can be written as
\begin{equation}
\label{trec_C}
t_{\rm rec}(z)\simeq \frac{5.88\times  10^{5}}{(1+z)^3} 
\frac{1}{C(z)}\,\, {\rm Myr}
\end{equation} 
\citep{MADAU}.

\begin{figure*}
 \includegraphics[width=0.45\textwidth]{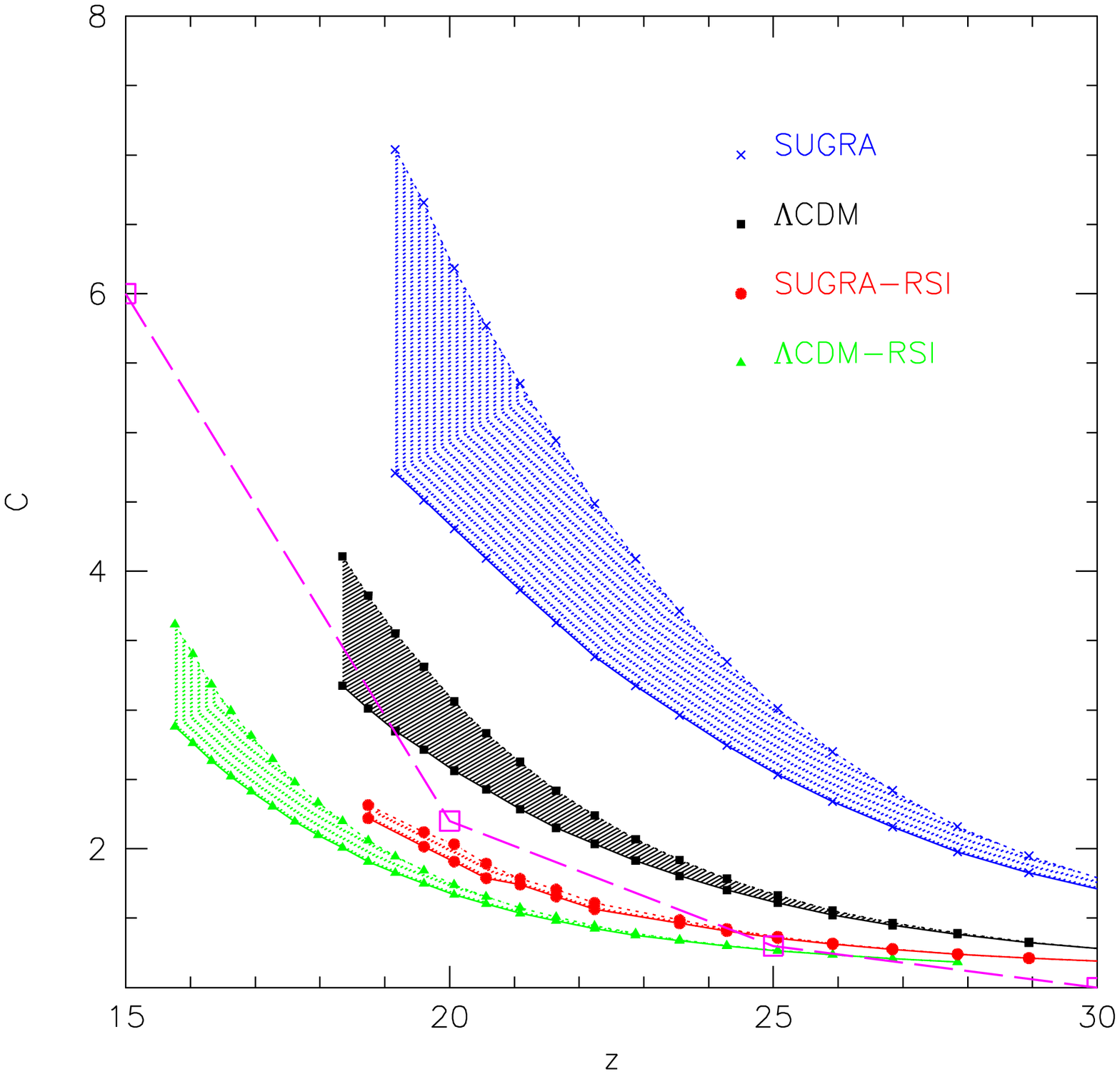} 
 \includegraphics[width=0.45\textwidth]{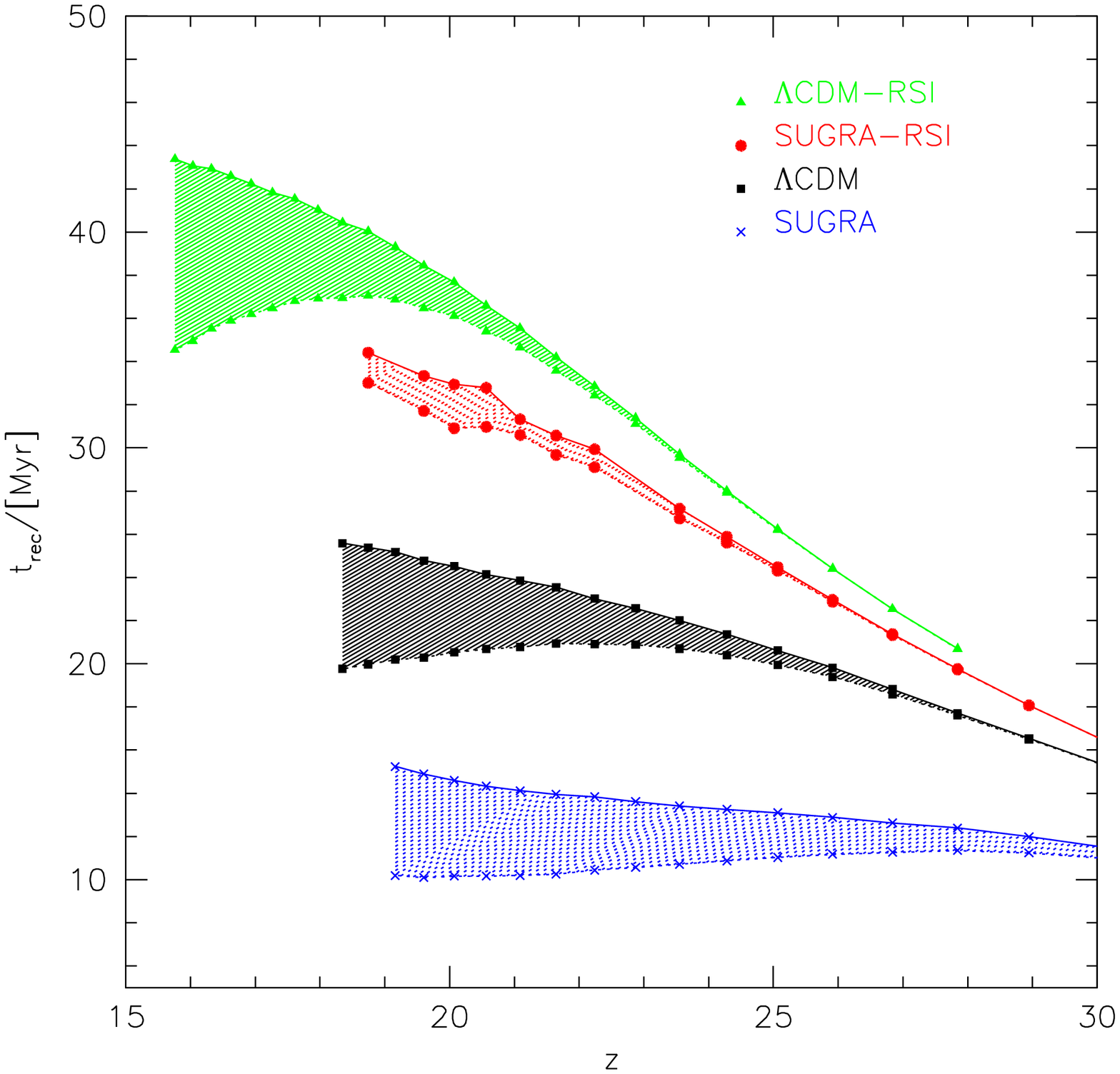}
\caption{\emph{Left panel}: The redshift evolution of the clumping
factor $C$ is shown for the different models (SUGRA, $\Lambda$CDM,
SUGRA-RSI and $\Lambda$CDM-RSI from top to bottom). The shaded regions
represent the uncertainty due to different choices for the maximum
overdensity (here assumed between 100, dotted lines, and 500, solid
lines; see the text for more detail). For reference, open squares and
the dashed line show the results obtained for the Z4 model in
\protect\cite{sfr2003}. The \emph{right panel} shows the evolution of
the recombination time $t_{\rm rec}$ in Myr for the same models.}
\label{fig:Clump}
\end{figure*}

The left panel in Fig.~\ref{fig:Clump} shows the redshift evolution of
the clumping factor $C$, which has been estimated in our SPH
simulations to be
\begin{equation}
C=\frac{\sum_i m_i\rho_i^{-1} \sum_j m_j \rho_j}{\left( 
\sum_k m_k  \right)^2}
\end{equation}
\citep{sfr2003}. To avoid numerical artifacts produced by gas
particles belonging to formed objects which would artificially
increase the value of $C$ due to their high density, the sum extends
over all gas particles whose density $\rho_i$ is smaller than a given
threshold.  In order to bracket realistic values for the threshold, we
show the results (by shaded regions) assuming a maximum overdensity in
the range between 100 and 500. We tested that the following analysis
depends only slightly on the exact threshold.

In general the growth of $C$ with time is quite regular for all
models.  For this reason we decided to fit its behaviour assuming a
parabolic relation, $C=a_0+a_1 z + a_2 z^2$. The resulting
best-fitting parameters $a_i$, obtained using for $C$ the
central values of the shaded regions with the corresponding dispersion,
are reported in Table \ref{tab:clump}. Notice that the relation for
the $\Lambda$CDM-RSI model is valid only for $z<22$, because of the
strong flattening at higher redshifts.  In line with our previous
results, we find that gas clumping decreases from SUGRA through
$\Lambda$CDM, SUGRA-RSI and $\Lambda$CDM-RSI. The dominant effect is
caused by the shape of primordial power spectrum. In fact, the
differences between the two RSI models are very small at all
redshifts, where $C$ is always between unity (corresponding to an
almost perfectly smooth gas distribution), and three. For the
$\Lambda$CDM model instead, the clumping factor starts to
significantly deviate from unity at $z\simeq25$, reaching four at
$z\sim 18$. As mentioned before, the gas distribution appears much
more clumpy in SUGRA, where $C$ starts to grow already from
$z\simeq30$, reaching $C\sim6$ at $z\simeq19$. Notice that our values
for $C$ in $\Lambda$CDM are larger by 50 per cent compared to those
obtained for the same model by
\cite{sfr2003} (see the open squares and the dashed line in the
plot). This discrepancy is caused both by the higher resolution of our
simulation and by the different treatment of the gas component in the
numerical code. Even if star formation is not included like in
\cite{sfr2003}, the fluid quantities (density, temperature and
chemical abundances) are accurately and self-consistently evolved in
our simulations, in particular also the cooling by molecular
hydrogen. Thus, they are expected to produce a more realistic gas
distribution in the diffuse intergalactic medium.

\begin{table}
\begin{center}
%\vspace{0.8cm}
\caption{The fitting parameters for the redshift evolution
of the clumping factor $C$ computed in the different simulations
(listed in Column 1). We adopt a parabolic relation, given by
$C=a_0+a_1 z + a_2 z^2$.
}
\begin{tabular}{lccc}
\hline
\hline
Model & $a_0$ & $a_1$ & $a_2$\\
\hline
SUGRA            & 30.379 & -1.858 & 0.030 \\
$\Lambda$CDM     & 17.295 & -1.093 & 0.019 \\
SUGRA-RSI        &  9.592 & -0.585 & 0.010 \\
$\Lambda$CDM-RSI & 20.086 & -1.655 & 0.037 \\
\hline
\hline
\label{tab:clump}
\end{tabular}
\end{center}

\vspace{0.8cm}
\end{table}

Knowing the clumpiness factor $C$, it is easy to estimate the
recombination time $t_{\rm rec}$ from Eq.~(\ref{trec_C}). The
corresponding results are plotted in the right panel of
Fig.~\ref{fig:Clump}. Of course, the maximum is obtained for the model
with the lowest $C$, i.e.~$\Lambda$CDM-RSI, in which the recombination
process requires the long time of almost 40 Myr at $z\simeq18$. The
values for $t_{\rm rec}$ slightly decrease to about 30 Myr for
SUGRA-RSI and to 20 Myr for $\Lambda$CDM at the same redshift, and to
less than 10 Myr at $z\simeq19$ for SUGRA model. The differences in
both $C$ and $t_{\rm rec}$ are again spread over almost an order of
magnitude because they are integrated observables like $n_{\rm
clouds}$.

\section{Implications for reionisation}\label{sect:reionization}

It is interesting to draw consequences from our results
regarding the phenomenology of the overall reionisation process, 
checking in particular how these scenarios might be constrained by 
data on the reionised optical depth. Even if more accurate and
quantitative estimates would require detailed simulations including
radiative transfer, we can reliably discuss the
problem of reionisation based on preceding work.

We first need a relation between the reionisation epoch and the total
Thomson optical depth $\tau$, which can be derived from the CMB data.
To compare the WMAP results to the predictions for the cosmological
model here considered, we compute the redshift evolution of $\tau$,
adopting a simple model which assumes that complete reionisation
occurs instantaneously at a some redshift $z$. The differential
Thomson optical depth for complete ionisation is given by
\begin{equation}
\d\tau(z)=\frac{n(z)\, \sigma_T\, c}{(1+z)\,H(z)}\,\d z\ ,
\label{eq:tau}
\end{equation}
where
\begin{equation}
\sigma_T = \frac{8\pi}{3} \left(\frac{e^2}{m_\mathrm{e}c^2} \right)^2 
\approx 6.65\times10^{-25}\,{\rm cm}^2
\end{equation}
is the Thomson cross section, $c$ is the speed of light and $H(z)$ is
Hubble parameter at $z$. The mean electron number density is given in
terms of the ionisation fraction $x_\mathrm{e}(z)$ by
$n(z)=n(z=0)(1+z)^3 x_e(z)$. Notice that in the previous equations the
only dependence on the cosmological model appears in $H(z)$ through
the Hubble constant and the expansion rates. For this reason the
optical depth is completely independent on the spectral parameters
(such as the power-spectrum index $n$), and the RSI predictions agree
exactly with the corresponding $n=1$ model.

\begin{figure}
 \includegraphics[width=0.45\textwidth]{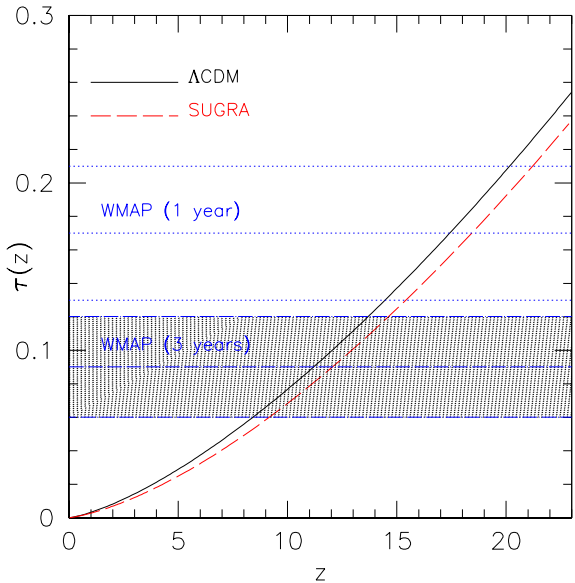}
\caption{The integrated Thompson optical depth $\tau$ for $\Lambda$CDM
(solid line) and SUGRA model with $w_0=-0.85$ (long-dashed line) are
compared to the WMAP constraints: first-year and three-years results
are shown by the regions within the dotted lines and by the shaded
region, respectively.}
\label{fig:tau}
\end{figure}

The integrated optical depth, computed assuming $x_e=1$ always, is
shown in Fig.~\ref{fig:tau} for both $\Lambda$CDM and SUGRA with
$w_0=-0.85$. The horizontal dotted lines represent the observational
estimate (with its 1-$\sigma$ error bars) obtained from the first-year
WMAP data \citep{kogut2003}, while the shaded region shows the more
recent values from the three-year WMAP data \citep{page2006}.\footnote
{We have to notice that our results for $\tau$ must be considered as
an extrapolation at redshifts higher than the reionisation epoch
suggested by WMAP: since $x_e$ vanishes, the curves should indeed
display a horizontal floor.}  When compared to standard $\Lambda$CDM,
SUGRA has a lower optical depth, with a difference of about
$\Delta\tau=0.02$ at $z\simeq17$ and $\Delta\tau=0.01$ at
$z\simeq10$. As expected, this implies an earlier reionisation epoch,
causing tension between the present model and the measured
reionisation optical depth \citep{spergel2006}, in addition to that
regarding other cosmological probes \citep[see, e.g.,
][]{spergel2003,knop2003,riess2004}.  Although a more accurate
estimate would require a realistic description of the reionisation
history, which in turn strongly depends on the free parameters of the
adopted modelling, we can then assume that the reionisation in both
models ($\Lambda$CDM and SUGRA) must occur and conclude at early
epoch, ranging between $z\simeq15$ and $z\simeq10$, to agree with the
measured optical depth within 1-$\sigma$. In general, these results
demonstrate that the records of the high-redshift expansion rate in
the physics of the primordial gas clouds may be used to constrain the
dark energy from reionisation data.  \cite{mainini2003} came to
similar conclusions without considering modifications to the physics
of gas clouds.

\begin{figure}
 \includegraphics[width=0.45\textwidth]{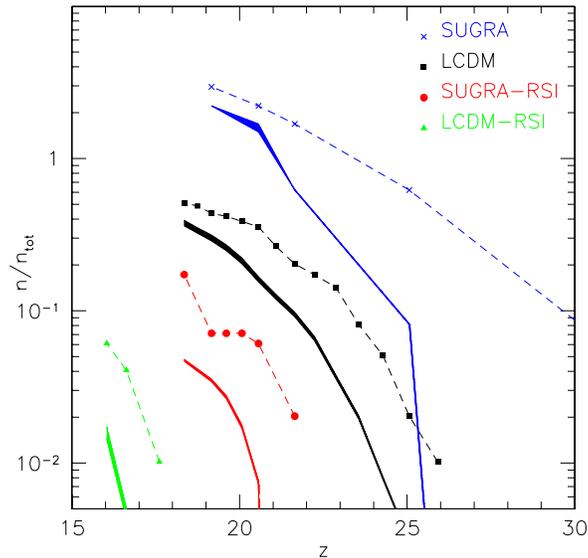}
\caption{
The redshift evolution of the number of ionizing photons $n_{\rm ion}$
(dashed lines) and recombined atoms (shaded regions corresponding to
the uncertainties for $t_{\rm rec}$ displayed in the right panel of
Fig.~\ref{fig:Clump}) are shown for the different models (SUGRA,
$\Lambda$CDM, SUGRA-RSI and $\Lambda$CDM-RSI from top to bottom).
Both quantities are normalized to the total number of atoms $n_{\rm
tot}$ (see more detail in the text).  }
\label{fig:qdot}
\end{figure}

We can now study in more detail how the reionisation process is going
on in the redshift range covered by our simulations.  First, we
compute the total number of ionising photons $n_{\rm ion}$. For this
goal, we can follow the usual `one star per halo' assumption for
mini-halos \citep[see, e.g.,][]{yoshida2003a,yoshida2003b} and safely
use the cloud number as a good approximation for the number of very
massive stars: in fact strong radiative feedback disfavours the
formation of multiple stars inside each primordial gas cloud.  Then we
set the mass of a Pop III star to be $500\,M_\odot$ and we use the
tables in \cite{schaerer2002} to estimate its lifetime ($\sim 1.9$
Myr) and the mean ionising flux along the evolutionary track ($\sim
6.8\times 10^{50}$ photons per second).  We also assume an optimistic
constant photon escape fraction of unity, which, however, can be
considered realistic when the gas distribution is reasonably smooth
\citep{oh2001}.  Detailed calculations by \cite{kitayama2004} 
indeed find that such a large escape fraction is plausible for small
mass haloes.  In order to complete the reionisation process, the
resulting values for $n_{\rm ion}$ must be at least as large as the
total number of hydrogen atoms in the simulation volume, which in our
case corresponds to $n_{\rm tot} \sim 4\times 10^{66}$.  In
Fig.~\ref{fig:qdot} the dashed lines show the redshift evolution for
$n_{\rm ion}$ for the different models. We notice that only the SUGRA
model is able to produce a sufficient number of ionising photons in
the redshift range considered by our simulations, while for all
remaining models the ratio $n_{\rm ion}/n_{\rm tot}$ is smaller than
unity.  In particular the two RSI models at the final simulation
outputs reach only the 10-15 per cent of the required photons.

Second, we have to consider that, as discussed before, recombination
is counteracting the reionisation process. Using the values for the
clumping factor and recombination time accurately derived from the
simulations (and shown in Fig.~\ref{fig:Clump}), we can easily compute
the cumulative number of recombined atoms, numerically solving
Eq.~(\ref{eq:dqdt}).  The results are also displayed in
Fig.~\ref{fig:qdot}, where the shaded regions correspond to the
uncertainties for $C$ and $t_{\rm rec}$ discussed in the previous
section.  The plot is suggesting that the recombination process is
rapid enough to significantly affect the fraction of ionised atoms. At
the final simulation outputs, we find that the ratio between
recombined and ionised atoms is about 75 per cent for SUGRA and
$\Lambda$CDM, and about 25 for the two RSI models.

Our analysis suggests that all models here considered, with the
exception of SUGRA, are producing an insufficient number of collapsed
objects to achieve complete reionisation in the redshift interval
covered by the simulations. This is true not only for both RSI models,
whose suppression of small scale power prevents the reionisation, but
also for the standard $\Lambda$CDM model.  These conclusions are in
qualitative agreement with the previous analysis performed by
\cite{yoshida2003b}, who, considering a simple model based on the
formation of first stars in mini-haloes by molecular cooling,
demonstrated that complete reionisation is reached when a density of
$50\sim100$ very massive stars per comoving $\mathrm{Mpc^3}$ is turned
on per $\sim1$ recombination time
\citep[see also][]{yoshida2003a,yoshida2003c}. 
We notice that for these models the results are not conflicting with
the three-year WMAP data which suggest a later reionisation epoch
compared to the first-year analysis. Allowing the simulations to
evolve up to $z\simeq10$ would certainly increase both the number of
gas clouds and the clumpiness, completing the reionisation process.

Finally, we notice that for the formation and evolution of cosmic
structures in SUGRA are so fast that the corresponding number of gas
clouds is so large and $t_{\rm rec}$ so low that a global reionisation
is expected at redshifts even higher than $z=20$.  As mentioned
already, this anticipation of the reionisation epoch could be in
conflict with observations \citep[see, e.g., the contour levels in
Fig. 3 of][]{spergel2006}, indicating the potential constraining power
of the dark energy records in the formation of primordial gas clouds
which we pointed out in this work.

\section{Conclusions}\label{sect:conclusions}

We presented hydrodynamic N-body simulations of the formation
of primordial gas clouds on scales of tens of kpc in a variety of
cosmological models, characterised by different dynamics in the
dark-energy component. Our main results are that the records of the
modified expansion rate are well evident in the population and the
clumpiness of such clouds. Cosmological models with the same
power-spectrum normalisation at present show earlier cloud formation
if the dynamics of the dark energy is enhanced, represented by an
equation of state parameter $w>-1$ as in quintessence models.

Within dark energy models compatible with the present data on cosmic
microwave background and large scale structure, the difference in the
integral population of clouds may vary by up to an order of magnitude,
as a consequence of the different differential efficiency for
structure formation. This is consistent with earlier results
indicating a higher concentration in dark matter haloes under similar
conditions \citep{dolag2004}.

Since abundance and clumpiness of structures are directly related to
the amount of primordial power on the corresponding scales, we varied
the shape of the primordial power spectrum by a running spectral index
reducing power on small scales within the confidence level of the
three-years WMAP data \citep{spergel2006}. As expected, we find that
the extra population and clumpiness of clouds produced by a higher
dark energy abundance compared with its level today might be mitigated
if the primordial spectral index is running, decreasing the power on
small scales.

Adopting a simple picture for the reionisation process, we derived
consequences for the reionisation itself, leading to an earlier
beginning  of the reionisation process in models where cloud
formation starts earlier. On the basis of these results, we are able
to identify possible tension between the WMAP data on the reionisation
optical depth and cosmological models whose dark energy is as
dynamical as in SUGRA quintessence models.

Our results demonstrate that the effects on cosmological structure
formation from a modified expansion history through different dark
energy models must be traced back to the formation of the first
clouds. In turn, this means that constraints on the dark energy, and
in particular its abundance at high redshifts, may be obtained by
forthcoming experiments aiming at measuring the abundance and the
clumpiness of primordial gas clouds.  In particular, the Atacama
Large Millimeter Array
(ALMA\footnote{www.eso.org/projects/alma/science/}) and the Mileura
Widefield Array
(MWA\footnote{www.haystack.mit.edu/ast/arrays/mwa/site/index.html})
will probe the early stages of structure formation, say between $z=6$
and $10$, where the reionization in progress should keep a record of
the population of reionizing primordial gas clouds.

\section*{acknowledgements}

Computations were performed on the IBM-SP4/5 at CINECA, Bologna, with
CPU time assigned under an INAF-CINECA grant, and the IBM-SP4 at the
computer centre of the Max Planck Society with CPU time assigned to
the Max Planck-Institut f\"ur Astrophysik. We are grateful to
C.~Gheller for his assistance. We acknowledge useful discussions with
Benedetta Ciardi, Bepi Tormen and Licia Verde.

\bibliographystyle{mn2e}
\bibliography{master2.bib}

\label{lastpage}
\end{document}